\documentstyle[12pt,epsfig]{article}
\newcommand{\bea}{\begin{eqnarray}}
\newcommand{\eea}{\end{eqnarray}}
\newcommand{\grgl}{\:\hbox to -0.2pt{\lower2.5pt\hbox{$\sim$}\hss}
           {\raise3pt\hbox{$>$}}\:}
\newcommand{\klgl}{\:\hbox to -0.2pt{\lower2.5pt\hbox{$\sim$}\hss}
           {\raise3pt\hbox{$<$}}\:}

\renewcommand{\Im}{{\mathcal{I}}}
\renewcommand{\Re}{{\mathcal{R}}}
\begin{document}
%
\renewcommand{\baselinestretch}{1}\normalsize
\vspace*{-2cm}
\hfill TUM-HEP-372/00

\vspace*{2cm}
\bigskip
\bigskip
\renewcommand{\baselinestretch}{1.4}\normalsize
\begin{center}
{\huge\bf{A consistent nonperturbative approach to thermal
damping-rates}}\Large{\footnote{Supported by the SFB 375 f\"ur Astroteilchenphysik der
Deutschen Forschungsgemeinschaft}} 
\end{center}
\bigskip
\renewcommand{\baselinestretch}{1}\normalsize
\begin{center}
{\Large{Bastian Bergerhoff}}\large{\footnote{bberger@physik.tu-muenchen.de}}
{\Large{ and J{\"u}rgen
Reingruber}}\large{\footnote{reingrub@physik.tu-muenchen.de}}\vspace*{0.3cm}\\
Institut f\"ur Theoretische Physik \\
Technische Universit\"at M\"unchen \\
James-Franck-Strasse, D-85748 Garching, Germany
\end{center}
\setcounter{footnote}{0}
\bigskip
\vspace*{3cm}\begin{abstract}
\noindent
We propose a nonperturbative scheme for the calculation of
thermal damping-rates using exact renormalization group (RG)-equations. 
Special emphasis is put on the thermal RG where first results for the rate were
given in 
\cite{Pietroni}.
We point out that in order to obtain a complete result that also 
reproduces the
known perturbative behaviour one has to take into account effects that were
neglected in  
\cite{Pietroni}.
We propose a well-defined way of doing the calculations that
reproduces perturbation theory in lowest order but goes considerably beyond
perturbative results and should be applicable also at second order
phase-transitions.
\end{abstract}
\bigskip
{\small{Pacs-No.s: 11.10.Wx,11.15.Tk,05.10.Cc,05.70.Ln}}

\newpage
\renewcommand{\baselinestretch}{1.1}\normalsize
Perturbation theory at finite temperatures is often invalidated by bad
infrared behaviour.
These problems can be solved by different resummation schemes.
One of the most powerful approaches to resummation is the
renormalization group of Wilson and others
\cite{WilsonRG}.
This
involves introducing an external scale and deriving functional differential
equations for the
dependence of generating functionals on this scale.
The right hand side of such an RG-equation can formally be interpreted
as a one-loop expression in the sense that 
it is of order $\hbar$
compared to the left hand side.
Nevertheless the Wilsonian RG constitutes a nonperturbative method and the
RG-equations are exact functional relations.
The approach is well known to correctly reproduce the infrared-behaviour of
theories even at second-order phase-transitions.

Even though the RG-equations define a nonperturbative approach, it is
possible to reproduce perturbation theory.
This is straightforward for one-loop results, but becomes rather
tedious if one goes to higher orders.
Quantities which are well described perturbatively for a range of temperatures
but for which perturbation theory fails e.g. at a phase-transition constitute a
major challenge for RG-equations.
This holds in particular for quantities where the one-loop contribution
vanishes.
The thermal damping-rate in $\varphi^4$-theory is an example for a quantity
where the lowest order perturbative contribution is two-loop.
If one aims at a reliable calculation of such quantities at all temperatures,
one has to reproduce two-loop perturbation theory as the leading term
{\it{and}} apply sensible nonperturbative resummations close to a possible
phase-transition.

A recent formulation of Wilsonian RG-equations in the Schwinger-Keldysh (CTP-)
formalism of real-time thermal field theory is particularily suited for
calculations of nonstatic thermal Green-functions
\cite{DAP1}.
It makes use of the fact that the propagators in the CTP-formalism
separate into the usual zero-temperature and a finite-T part.
The finite-T parts only contribute on-shell and depend only on the
three-momenta. 
The cutoff modifies the thermal part and
one introduces cut-off propagators of the form
\bea
D_\Lambda(k) &=& \left(
\begin{array}{cc} 
\Delta & 0 \\
0 & -\Delta^* \end{array} \right) + 
(\Delta - \Delta^*) 
\left(
\begin{array}{cc}
0 &
\Theta(-k_0) \\
\Theta(k_0)  &
0 \end{array} \right) + \nonumber \\
&& + (\Delta - \Delta^*) \Theta(|\vec{k}|,\Lambda) N(|k_0|)
\left(
\begin{array}{cc}
1 & 1 \\ 1 & 1 
 \end{array} \right)
\label{props}
\eea
where $\Theta(|\vec{k}|,\Lambda)$ is a possibly smeared out step function.
We will in the following use a sharp cutoff, i.e.~set
$\Theta(|\vec{k}|,\Lambda) = \Theta(|\vec{k}|-\Lambda)$.
Thus for finite $\Lambda$, the propagation of thermal modes with three-momemtum
small compared to $\Lambda$ (soft modes) is supressed, while the hard thermal
modes are unmodified.
Inserting this propagator into the usual expression for the generating
functional $Z[J]$ one obtains a $\Lambda$-dependent functional $Z_\Lambda$.
Introducing the modified Legendre-transform 
\bea
\Gamma_\Lambda[\Phi] = -i \ln Z_\Lambda[J] - J \cdot \Phi - \frac{1}{2} \Phi \cdot
\left( D_\Lambda \right)^{-1} \cdot \Phi
\label{GammaLambda}
\eea
one readily obtains for the scale-dependence of $\Gamma_\Lambda$ the exact
functional differential equation 
\cite{DAP1}
\bea
\Lambda \frac{\partial \Gamma_\Lambda[\Phi]}{\partial \Lambda} = \frac{i}{2}
{\mbox{Tr}} \Lambda \frac{\partial D_\Lambda^{-1}}{\partial \Lambda} \left(
D_\Lambda^{-1} + \frac{\delta^2 \Gamma_\Lambda[\Phi]}{\delta \Phi \delta \Phi}
\right)^{-1}
\label{TRG}
\eea
This is the {\it{thermal renormalization group equation}} (TRG) and will be the
starting point of our discussion.

Being a differential equation, (\ref{TRG}) of course has to be supplemented
by boundary conditions.
As discussed in 
\cite{DAP1},
in the limit $\Lambda \rightarrow \infty$, the effective action
$\Gamma_\Lambda$ is trivially obtained from the $(T\!=\!0)$-effective action of the
theory.
Also, in the limit $\Lambda \rightarrow 0$ the full finite temperature
CTP-effective action $\Gamma[\Phi]$ is obtained.

RG-equations for Green-functions are simply obtained from (\ref{TRG}) by
taking functional derivatives with respect to $\Phi$.
For the damping-rate, we are interested in the imaginary part of the two-point
function.
The flow-equation for the two-point function reads in terms of the thermal
fields $\varphi_i$ with $i=1,2$ ($\varphi_0$ is some background configuration)
\bea
\left. \Lambda \frac{\partial}{\partial \Lambda} \frac{\delta^2
\Gamma_\Lambda}{\delta \varphi_i \delta \varphi_j} \right|_{\varphi_0} &=&
\left. \frac{1}{2} {\mbox{Tr}} K_\Lambda \frac{\delta^4 \Gamma_\Lambda}{\delta
\varphi_i \delta \varphi_j \delta \Phi \delta \Phi} \right|_{\varphi_0} -
\nonumber \\
&& \quad  - \left( \left. \frac{1}{2} {\mbox{Tr}} K_\Lambda \frac{\delta^3
\Gamma_\Lambda}{\delta \varphi_i \delta \Phi \delta \Phi} {\mathcal{D}}
\frac{\delta^3 \Gamma_\Lambda}{\delta \varphi_j \delta \Phi \delta \Phi}
\right|_{\varphi_0} + (i \leftrightarrow j) \right)
\label{FlowTPF}
\eea
where $K_\Lambda = -i {\mathcal{D}} \Lambda \frac{\partial
D_\Lambda^{-1}}{\partial \Lambda} {\mathcal{D}} $ is the kernel of the TRG
and ${\mathcal{D}} = \left( D_\Lambda^{-1} + \Gamma_\Lambda^{(2)}
\right)^{-1}$ is the full two-point function
\cite{DAP1}. 
Even though in this representation the flow-equation appears like a plain one-loop equation,
due to the fact that all $n$-point functions involved are {\it{full}} $n$-point
functions it is actually an exact, nonperturbative expression.

Let us then assume that we want to calculate the imaginary part of the thermal
two-point function in a scalar theory with
unbroken $Z_2$-symmetry.
Perturbatively in this case the lowest contribution to
the imaginary part of the self-energy occurs on two-loop level. 
As out in 
\cite{Pietroni},
in such a situation the contribution to the imaginary part of (\ref{FlowTPF})
has to come from an imaginary part of the full four-point function
(by the
$Z_2$-symmetry, the second contribution in (\ref{FlowTPF}) vanishes
identically for all $\Lambda$).
We thus have to solve a coupled system of at least two flow-equations,
(\ref{FlowTPF}) and a corresponding equation for the imaginary part of the
four-point function.

First however let us make
some remarks concerning the thermal indice which appear in real-time
formulations of thermal field-theory 
\cite{LeBellac}.
The physical fields always carry a thermal index 1 and one
is in general interested in the calculation of Green-functions for those
fields.
Since the propagators are nondiagonal in the thermal indice, 
in the calculation of higher
loop contributions one has to include vertice of the 2-fields
nevertheless -- neglecting those contributions yields singular expressions.
If we use the TRG to compute Green-functions, we have to use full vertice
according to (\ref{TRG}).
In this case, one has to allow for vertice with mixed thermal
structure. 

Furthermore, what enters the physical self-energy (and thus the damping-rate)
is not the Green-function $\frac{\delta^2 \Gamma}{\delta \varphi_1 \delta
\varphi_1}$, but rather a specific combination of
the $\frac{\delta^2 \Gamma}{\delta \varphi_i \delta \varphi_j}$
\cite{LeBellac},
known as $\bar{\Pi}(p)$ in the literature.
Assuming a Schwinger-Dyson equation for the full propagator it is easy to show
that for the real- and imaginary parts of $\bar{\Pi}$ the following relations
hold:
\bea
\Re \bar{\Pi}(p) &=& \Re \frac{\delta^2 \Gamma}{\delta \varphi_1(p) \delta
\varphi_1(-p)} \nonumber \\
\Im \bar{\Pi}(p) &=& \epsilon(p_0) \Im \left( \frac{\delta^2 \Gamma}{\delta
\varphi_1(p) \delta \varphi_1(-p)} +\frac{\delta^2 \Gamma}{\delta \varphi_1(p)
\delta \varphi_2(-p)} \right)
\label{Pibar}
\eea
($\epsilon(p_0)$ is the sign function). 
$\Im \bar{\Pi}$ can also be obtained as
\bea
\Im \bar{\Pi}(p) &=& \frac{1}{1+2 N(|p_0|)} \Im \frac{\delta^2 \Gamma}{\delta
\varphi_1(p) \delta \varphi_1(-p)}
\label{Pibar2}
\eea
Note that this expression explicitely involves a distribution function, whereas
the second equation of (\ref{Pibar}) does not.
We will below use (\ref{Pibar}) for a calculation of the physical 
self-energy.\footnote{In \cite{Pietroni}, (\ref{Pibar2}) was used for the
calculation of the rate. 
We believe for a number of reasons that will become clear below that it is
preferrable to use (\ref{Pibar}).
Nevertheless, in principle for an exact solution of the TRG, the outcome should
be identical.}
Another important
feature of real-time thermal field-theories is that the effective action
has the following symmetry
\cite{NiemiSemenoff}
\bea
\Gamma[\varphi_1, \varphi_2] = - \Gamma^*[\varphi_2^*, \varphi_1^*]
\label{Z2symm}
\eea
For real, momentum independent vertice (in the present theory with a real
scalar field in configuration space), this gives the following relations:
\bea
\Gamma^{(n)}_{i_1 i_2 ... i_n} = - \Gamma^{(n)}_{\bar{i}_1 \bar{i}_2
... \bar{i}_n}
\label{Z2forVerts}
\eea
where $\bar{i} = 2$ for $i = 1$ and vice versa.
Furthermore one may introduce a functional $\bar{\Gamma}[\phi]$ by
\bea
\frac{\delta \bar{\Gamma}[\phi]}{\delta \phi} = \left. \frac{\delta
\Gamma}{\delta \varphi_1} \right|_{\varphi_1 = \phi, \varphi_2 =
\varphi_2[\phi]}
\label{Gammabar}
\eea
where $\varphi_2[\phi]$ is the solution of the field-equation for
$\varphi_2$.
For momentum-independent vertice, it is easy to see that one has
\bea
\bar{\Gamma}^{(3)} = \sum_{i,j=1,2} \left. \Gamma^{(3)}_{1ij}
\right|_{\varphi_a = \phi} \quad ; \quad 
\bar{\Gamma}^{(4)} = \sum_{i,j,k=1,2} \left. \Gamma^{(4)}_{1ijk}
\right|_{\varphi_a = \phi} 
\label{barVerts}
\eea
and so on. 
These relations trivially generalize to $\Gamma_\Lambda$ and we will make use
of them below.

Let us now turn to the calculation of the imaginary part of the self-energy 
of (at $T\!=\!0$) massless $\varphi^4$-theory.
In order to obtain an equation for the scale-dependence of $\Im \bar{\Pi}$, we
plug the flow-equations for the two-point functions with 1-1 and 1-2
external legs into the second relation in (\ref{Pibar}).
We now make a crucial approximation: We neglect the imaginary part of the
self-energy on the right hand side of all flow-equations (for the
present work we also assume $ \Re \bar{\Pi}$ to be momentum-independent,
which is however not crucial for the results discussed and could be
straightforwardly improved).
This may be viewed as a "quasiparticle-approximation". 
As we will see below, it does not influence the leading behaviour for small
couplings.
Making this approximation however means that we will not be resumming imaginary
parts in the calculations -- a fact important to keep in mind.

In this case, the kernel takes the form
\cite{DAP1}
\bea
K_\Lambda(k) &=& 
- 2 \pi \delta(k^2-m_\Lambda^2) \Lambda 
\delta(|\vec{k}|-\Lambda) N(|k_0|) \left( \begin{array}{cc}
1 & 1 \\ 1 & 1 \end{array}\right) 
\label{kernel}
\eea
($m_\Lambda$ is the real part of the self-energy) and is purely real.
It is thus immediatly clear that all contributions to the flow of $\Im
\bar{\Pi}$ are from the imaginary part of the full four-point function.
Also, due to the specific combination of $\Gamma_{\Lambda,11}^{(2)}$ and 
$\Gamma_{\Lambda,12}^{(2)}$ appearing in (\ref{Pibar}), taking the trace over
thermal indice in the flow-equation for $\bar{\Pi}$ leaves us with the
following result:
\bea
\Lambda \frac{\partial \Im \bar{\Pi}_\Lambda\!(p)}{\partial \Lambda} = 
- \frac{\Lambda^3}{4\pi^2} \frac{N(\omega_\Lambda)}{\omega_\Lambda} 
\epsilon(p_0) \!\!\! \sum_{i,j,k=1,2} \!\!\! 
\Im \Gamma_{\Lambda,1ijk}^{(4)}(p,\!-p,Q_\Lambda,\!-Q_\Lambda) 
\label{flow1}
\eea
In (\ref{flow1}) we use the notation
\bea
\omega_\Lambda = \sqrt{\Lambda^2+m_\Lambda^2} \quad , \quad Q_\Lambda =
(\omega_\Lambda, |\vec{Q_\Lambda}| = \Lambda)
\label{notLambda}
\eea
For the quantities appearing on the right hand side of (\ref{flow1}), we again
need the RG-equations governing their scale-dependence.
Let us first discuss the flow-equation for the coupling appearing in
(\ref{flow1}).
We use the notation
\bea
\Im \!\!\!\! \sum_{i,j,k=1,2} \!\!\!
\Gamma_{\Lambda,1ijk}^{(4)}(p,-p,q,-q) = \Im \bar{\Gamma}_{\Lambda}^{(4)}
(p,-p,q,-q)
\label{img4}
\eea
which is suggested by (\ref{barVerts}).
We now make a further approximation, which is on equal footing with the
quasiparticle-approximation done above: 
We neglect the imaginary parts of all couplings on the right hand side of the
flow-equation for $\Im \bar{\Gamma}^{(4)}_\Lambda$.
This should of course not be done in (\ref{flow1}), since there one would
disregard the leading (and in fact only) contribution. 
Within this approximation we find:
\bea
\Lambda \frac{\partial}{\partial \Lambda} \Im
\bar{\Gamma}_\Lambda^{(4)}(p,-p,q,-q)\!\!\! &=&\!\!\! - \frac{\Lambda}{2}
\left(\bar{\Gamma}^{(4)}_\Lambda\right)^2 
\int \!\! \frac{d^4 l}{(2\pi)^2}
\delta(l^2\!-\!m_\Lambda^2) \theta(|\vec{l}|^2\!+\!m_\Lambda^2) \delta
(|\vec{l}|\!-\!\Lambda) N(|l_0|)\!\! \times \nonumber \\
&& \qquad \qquad \qquad \times \sum_{Q=\pm q} \delta((l+p+Q)^2-m_\Lambda^2) \epsilon(l_0+p_0+Q_0) 
\label{flow2}
\eea
Note that (\ref{flow2}) is a completely well-defined expression.
Furthermore the vertex appearing here is just the combination defined in
(\ref{barVerts}), and we have taken into account vertice with both 1- and
2-external legs.
This is only possible for the specific sum in (\ref{img4}).
Neglecting vertice with 2-legs within the present approach would
yield ill-defined products of distributions.
We have checked that the inclusion of trilinear couplings in the present
approximations (i.e. considering a theory with broken symmetry at low
temperatures) is possible along the same line of arguments.
Thus the results depicted in (\ref{flow1}) and (\ref{flow2}), together with a
flow-equation for the real part of the effective action at constant fields,
constitute a well-defined nonperturbative method for the calculation of
thermal rates.
This is the first important point in the present work.

Even though the approach using Wilsonian RG-equations is nonperturbative
one wants to recover the leading perturbative results in
situations where perturbation theory is valid.
This is a nontrivial problem, since $\Im \bar{\Pi}$  in the theory under study
vanishes to one-loop order.

The loop expansion can be reconstructed from
Wilsonian RG-equations iteratively, making use of the fact that the right hand
side of (\ref{TRG}) is down by a factor of $\hbar$ compared to the left hand
side.
In order to obtain a result for $\Im \bar{\Pi}$ to order $\hbar^2$, we need
the right hand side of (\ref{flow1}) to order $\hbar$.
The quantities appearing on the right hand side of (\ref{flow1}) are the
thermal mass $m_\Lambda$ and the imaginary part of the
four-point function at finite $\Lambda$.
This imaginary part is a pure quantum effect and
thus is itself of order $\hbar$, so we need not consider corrections to the
mass.

In order to find the imaginary part of the four-point function (which has to be
calculated for the specific choice of momenta entering in (\ref{flow1}) and at
nonvanishing external scale $\Lambda$), we turn to the corresponding
flow-equation (\ref{flow2}).
To order $\hbar$ we may neglect all
loop corrections to the quantities appearing on the right hand side of
(\ref{flow2}).
Thus to this order the couplings and masses are $\Lambda$-independent and
the integration of (\ref{flow2}) may be performed
noting that $-\delta(|\vec{l}|-\Lambda) = \partial_\Lambda
\theta(|\vec{l}|-\Lambda)$.
One finds
\bea
& &\Im \bar{\Gamma}_\Lambda^{(4)}(p,-p,q,-q) = \Im
\bar{\Gamma}_{\Lambda=\infty}^{(4)}(p,-p,q,-q) + \nonumber \\
&& \quad + \frac{1}{2}
\left(\bar{\Gamma}^{(4)} \right)^2 \!\! \int \!\! \frac{d^4 l}{(2\pi)^2}
\delta(l^2\!-\!m^2) \theta(|\vec{l}|^2\!+\!m^2) \theta (|\vec{l}|\!-\!\Lambda) N(|l_0|)\!\! \times \nonumber \\
&& \qquad \quad \times \sum_{Q=\pm q} \delta((l+p+Q)^2-m^2) \epsilon(l_0+p_0+Q_0) 
\label{Iflow2}
\eea
Plugging this result into the right hand side of the flow-equation for $\Im
\bar{\Pi}$ and performing the remaining $\Lambda$-integration should then give
the desired imaginary part of the self energy to order $\hbar^2$.

The result of a perturbative calculation may be taken e.g.~from
\cite{WangHeinz}
where one finds
\bea
& &\Im \bar{\Pi}(p) = - \pi \epsilon(p_0)
\frac{\left(\bar{\Gamma}^{(4)} \right)^2}{6} 
\int \frac{d^3 \vec{k}}{(2\pi)^3} \frac{d^3 \vec{q}}{(2\pi)^3} \frac{1}{8
\omega_k \omega_q \omega_r} \times \nonumber \\
&& \qquad \times \biggl \lbrace \left[ 
\delta(p_0 \! + \! \omega_k \! + \! \omega_q \! + \! \omega_r)-
\delta(p_0 \! - \! \omega_k \! - \! \omega_q \! - \! \omega_r)\right] \times \nonumber \\
&& \qquad \qquad \qquad \qquad \times \left[ 1 \! + \! N_k \! +\! N_q \! +\! N_r\!  +\!  N_k N_q \! +\!  N_k N_r \! +\!  N_q N_r  \right] +
\nonumber \\
&& \qquad \qquad + \left[ 
\delta(p_0 \! +\!  \omega_k \! +\!  \omega_q \! -\!  \omega_r)-
\delta(p_0 \! -\!  \omega_k \! -\!  \omega_q \! +\!  \omega_r)\right] 
\left[ N_r \! +\!  N_k N_r \! +\!  N_q N_r  \! -\!  N_k N_q  \right] +
\nonumber \\
&& \qquad \qquad + \left[ 
\delta(p_0 \! +\!  \omega_k \! -\!  \omega_q \! +\!  \omega_r)-
\delta(p_0 \! -\!  \omega_k \! +\!  \omega_q \! -\!  \omega_r)\right] 
\left[ N_q \! +\!  N_k N_q \! +\!  N_r N_q  \! -\!  N_k N_r \right] +
\nonumber \\
&& \qquad \qquad + \left[ 
\delta(p_0 \! -\!  \omega_k \! +\!  \omega_q \! +\!  \omega_r)-
\delta(p_0 \! +\!  \omega_k \! -\!  \omega_q \! -\!  \omega_r)\right] 
\left[ N_k \! +\!  N_k N_q \! +\!  N_k N_r  \! -\!  N_q N_r  \right] 
\biggr \rbrace
\label{twoloop}
\eea
where again $\omega_p = \sqrt{\vec{p}^2+m^2}$ and $\vec{r} =
\vec{k}+\vec{q}+\vec{p}$.
$N_p$ is the Bose-distribution with energy $\omega_p$.
The part $\propto 1$ is of course the $(T\!=\!0)$-imaginary part. 
In the present method this part enters via the boundary conditions and is not
calculable within the TRG.
The other parts however have to come out of our result from (\ref{flow1}) and
(\ref{flow2}).

An important observation at this point is that for the one-loop contribution
in (\ref{Iflow2}) one has two parts, one being the boundary part at
$\Lambda\!=\!\infty$ which is given by the one-loop $(T\!=\!0)$-result and the second one
being proportional to $N$ and representing the contribution from thermal
fluctuations.
If we simply start with the tree-level effective action at $T\!=\!0$ -- that is
set the imaginary part of the four point function to 0 for $\Lambda \rightarrow
\infty$ -- and plug the remaining contribution from (\ref{Iflow2}) into
(\ref{flow1}), we only reproduce the contributions to $\Im \bar{\Pi}$ which are
bilinear in the distribution functions.
Dropping all $(T\!=\!0)$-quantum contributions for the boundary value of the flow of
$n$-point functions is however routinely done in applications of the TRG
(\cite{Pietroni,DAP1,B1,BJ1,BJJ}).
Since the thermal damping-rate vanishes to one-loop,
one does not reproduce the leading order 
perturbative result (\ref{twoloop}) and will not be able to give
quantitatively reliable results for all temperatures.

To end the discussion of the perturbative result, we thus need the $\Lambda
\rightarrow \infty$ or $(T\!=\!0)$-value of the imaginary part of the four-point
function to one loop.
This is easily found to be
\bea
\Im \bar{\Gamma}_{\Lambda=\infty}^{(4)}(p,-p,q,-q) \!\!&=&\!\! 
\sum_{Q=\pm q} \! \frac{\pi}{8} \left(\bar{\Gamma}_{T=0}^{(4)}\right)^2 
\!\int \!\!\frac{d^3 \vec{k}}{(2\pi)^3} \frac{1}{\omega_k \omega_{k+p+Q}}
\times \nonumber \\
&&\quad \times \left[ \delta (p_0\! +\! Q_0\! -\! \omega_k\! -\! \omega_{k+p+Q}) -
\delta(p_0\! +\! Q_0\! +\! \omega_k\! +\! \omega_{k+p+Q}) \right]
\label{img4t0}
\eea
Using this result in (\ref{Iflow2}) and plugging the resulting expression for
$\Im \bar{\Gamma}_{\Lambda}^{(4)}(p,-p,Q_\Lambda,-Q_\Lambda)$ into the
flow-equation for $\Im \bar{\Pi}(p)$ given in (\ref{flow1}) one may indeed do
the integration with respect to $\Lambda$ to find for $\Lambda=0$ the
perturbative result (\ref{twoloop}) up to its $(T\!=\!0)$-part.\footnote{In order
to obtain the explicit form given in (\ref{twoloop}), symmetrization in the
three-momenta is necessary. The calculation is however straightforward.}

In order to go beyond the leading order two-loop result, 
we will take into account the thermal
renormalization of the real parts of both the self-energy as well as the
momentum-independent parts of the vertice.
We perform a derivative expansion of the full effective action
$\bar{\Gamma}_\Lambda$ and approximate it by the effective potential and a
standard kinetic term.
This approximation is discussed in detail in 
\cite{DAP1,B1,BJ1,BJJ} 
where also the resulting flow-equations may be found.
We solve the flow-equation for the effective potential as discussed in 
\cite{B1,BJ1,BJJ}
and thus resum the momentum-independent (thermal) parts of $n$-point 
functions with an arbitrary number of external legs -- this resummation
goes considerably beyond the usual daisy- or superdaisy-resummations.
It can be systematically improved by relaxing the 
derivative-approximation.

As the boundary condition for the flow of the effective potential we
will use the tree-level effective potential of a massless $\varphi^4$-theory,
$U_{\Lambda=\infty} = \frac{g_{T=0}}{24} \phi^2$.
This amounts to neglecting all $(T\!=\!0)$-loop corrections
to the effective potential.
Together with the boundary conditions and flow-equations for the imaginary
parts as discussed above we thus have a closed system of differential equations\footnote{Note that for the solution of
the flow-equation for $\Im \bar{\Gamma}_\Lambda^{(4)}(p,-p,q,-q)$, the external momentum $q$
is not given by (\ref{notLambda}) with the running values of $\Lambda$ and
$\omega_\Lambda$. 
Instead it has to be considered fixed for some independent combination $(\Lambda',\omega_{\Lambda'})$.}
which may be solved numerically.

The resulting imaginary part of the self-energy is connected to the thermal
damping-rate $\gamma(T)$ through
\bea
\gamma(T) = \frac{\Im \bar{\Pi}_{\Lambda=0}(p_0=m_T,\vec{p}=0)}{2 m_T}
\label{gamma}
\eea
Perturbatively, using daisy-resummed perturbation theory one finds for a 
massless theory
\cite{WangHeinz,Parwani,Jeon}
\bea
\gamma_{pert}(T) = \frac{g_{T=0}^{3/2}}{64 \sqrt{24} \pi} \sqrt{1-\frac{3
g_{T=0}^{1/2}}{\sqrt{24}\pi }} T
\label{gammapert}
\eea
In figure 1 we display the ratio of the result obtained using the
flow-equations and (\ref{gammapert}) as a
function of $g_{T=0}$.
Indeed the TRG reproduces daisy-resummed perturbation theory for
$g_{T=0} \rightarrow 0$, even though the leading perturbative result is
two-loop.
For finite values of the coupling, the different resummations yield different
results.
Let us again point out that the result of the renormalization
group-calculations performed here can be understood as a fully resummed result
in leading order in an expansion in the anomalous dimension and the
quasiparticle approximation. 
It goes beyond the 
resummations used in the literature (it in particular trivially includes the
daisy- and superdaisy-schemes
\cite{DAP1}).

Figure 2 shows the dependence of the real- and imaginary parts of the
self-energy on the external scale $\Lambda$ at fixed $g_{T=0}$.
We note the typical behaviour: Small corrections to the $(T\!=\!0)$-values for
large $\Lambda/T$ due to Boltzmann-suppression of the thermal fluctuations
and saturation of the values for $\Lambda \ll m_\Lambda$.
The limit $\Lambda \rightarrow 0$ is completely safe.
\hspace*{0.3cm}
\begin{figure}
\centering\epsfig{file=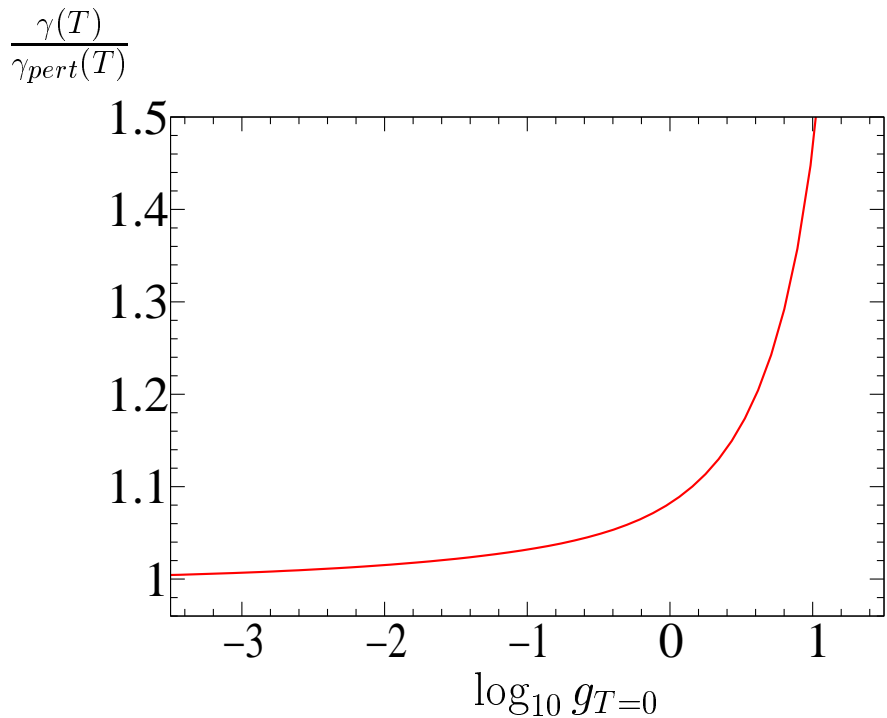,width=12cm,clip=}
\caption{$\gamma(T)/\gamma_{pert}(T)$ as function of $\log g_{T=0}$.}
\end{figure}
\hspace*{0.5cm}
\begin{figure}
\centering\epsfig{file=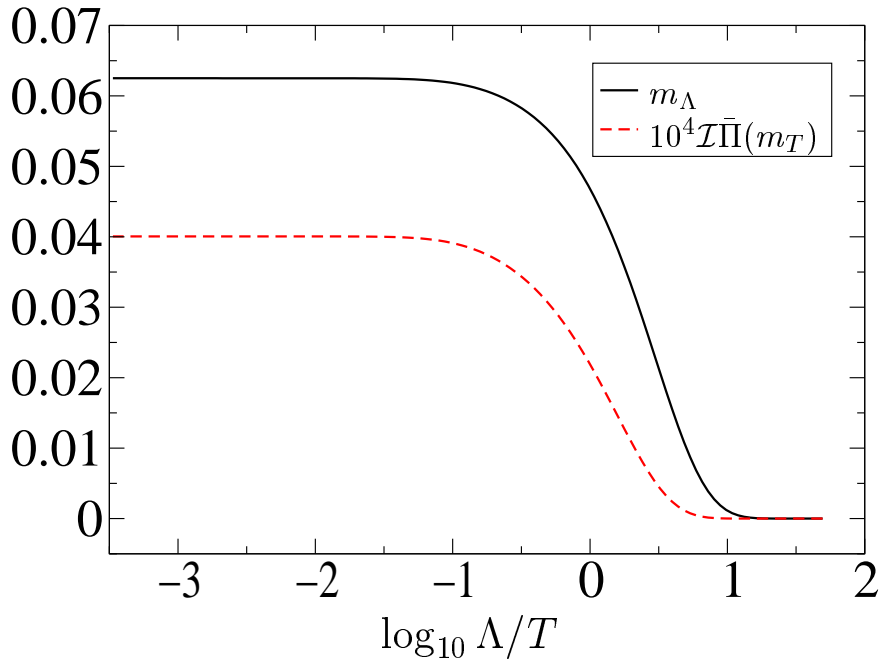,width=12cm,clip=}
\caption{The thermal mass $m_\Lambda$ and the imaginary part of the self energy
$\Im \bar{\Pi}_\Lambda(m_T)$ as functions of $\log \Lambda/T$. $g_{T=0} = 0.1$ was
chosen and dimensionful quantities are in units of $T$.}
\end{figure}

We close this letter by commenting on one of the most important qualitative
points made in 
\cite{Pietroni},
namely that a calculation of the damping-rate from the TRG reproduces critical
slowing down, i.e. the fact that the rate vanishes as the critical temperature
of a theory with (at $T\!=\!0$) spontaneously broken $Z_2$-symmetry is approached.
We believe that even though this conclusion was reached on the basis of an
incomplete calculation it still holds if one uses the
consistent approach laid out in the present paper.
The reason why this should be the case is easy to see: 
The rate is an on-shell quantity with vanishing external momentum
and as the critical temperature of the (second-order)
phase-transition is approached, the external energy thus vanishes.
On the other hand, in the framework of Wilsonian RG-equations internal
propagators are never massless for any $\Lambda > 0$.
Close to the phase-transition, the masses and coupling exhibit
three-dimensional scaling behaviour, i.e. the renormalized mass and coupling
vanish as
\bea
m_T \propto (T-T_c)^\nu \quad ; \quad g_T \propto (T-T_c)^\nu
\label{scaling}
\eea
In such a case, the limit $m_T \rightarrow 0$ should be completely regular
and the scaling arguments also given in 
\cite{Pietroni}
should still hold:
The perturbative result for the imaginary part of the self-energy behaves for
small $m_T$ as $\bar{\Pi}(m_T,0) \sim g_{T=0}^2 \ln m_T$
and thus one has for the rate
\bea
\gamma(T) \sim \frac{g_{T=0}^2}{m_T} \ln m_T
\label{ratesmallm}
\eea
Consistent resummation using the Wilsonian RG replaces the coupling $g_{T=0}$
with the renormalized coupling $g_T$ and one obtains with $t=T-T_c$
\bea
\gamma(T\rightarrow T_c) \propto t^\nu \ln t \rightarrow 0
\label{csd}
\eea
for positive $\nu$ (in the present model, $\nu$ is found to be $\sim 0.63$
\cite{ZinnJustin,BJ1,BJJ}).

Within the scheme presented here we now have for the first time a complete
approach that allows us to study linear-response and static
quantities in a nonperturbative manner also in the critical regime,
while reproducing the perturbative results where they are valid.
As shown in 
\cite{BJJ},
the TRG can also be formulated for theories involving fermionic degrees of
freedom.
It should be very interesting to apply this nonperturbative method e.g.~to questions
related to the physics of nuclear matter as the chiral phase-transition is
approached.

\bigskip
{\bf{Acknowledgements:}} We thank M. Pietroni for helpful discussions on his results.


\begin{thebibliography}{99}
\bibitem{Pietroni} M. Pietroni, Phys. Rev. Lett. {\bf{81}} (1998), 2424.
\bibitem{WilsonRG} K.G. Wilson and J.G. Kogut, Phys. Rep. {\bf{12}} (1974) 75.
\bibitem{DAP1} M. D'Attanasio and M. Pietroni, Nucl. Phys. {\bf{B472}} (1996)
711.
\bibitem{LeBellac} M. Le Bellac, "Thermal Field Theory" (Cambridge 1996).
\bibitem{NiemiSemenoff} A. Niemi and G. Semenoff, Ann. of Phys. {\bf{152}}
(1984) 105.
\bibitem{WangHeinz} E. Wang and U. Heinz, Phys. Rev. {\bf{D53}} (1996), 899.
\bibitem{B1} B. Bergerhoff, Phys. Lett. {\bf{B437}} (1998), 381.
\bibitem{BJ1} B. Bergerhoff and J. Reingruber, Phys. Rev. {\bf{D60}} (1999),
105036.
\bibitem{BJJ} B. Bergerhoff, J. Manus and J. Reingruber, hep-ph/9912474, to
appear in Phys. Rev. {\bf{D}}.
\bibitem{Parwani} R.R. Parwani, Phys. Rev. {\bf{D45}} (1992), 4695, {\bf{D48}}
(1993), 5965 (E).
\bibitem{Jeon} S. Jeon, Phys. Rev. {\bf{D52}} (1995), 3591.
\bibitem{ZinnJustin} J. Zinn-Justin, "Quantum Field Theory and Critical
Phenomena" (Oxford 1989).
\end{thebibliography}
\end{document}